\documentclass[a4paper,11pt]{article}
\usepackage{pos}
\usepackage{CJK}
\usepackage{amsmath}
\DeclareMathOperator{\Tr}{Tr}

\title{Signal-to-noise improvement through neural network contour deformations for 3D $SU(2)$ lattice gauge theory}
\ShortTitle{Neural network contour deformations for lattice gauge theory}

\author[a,b]{William~Detmold}
\author[c]{Gurtej~Kanwar}
\author*[a,b]{Yin~Lin}
\author[a,b]{Phiala~E.~Shanahan}
\author[d]{Michael~L.~Wagman}

\affiliation[a]{Center for Theoretical Physics, Massachusetts Institute of Technology, Cambridge, MA 02139, USA}
\affiliation[b]{The NSF AI Institute for Artificial Intelligence and Fundamental Interactions}
\affiliation[c]{Albert Einstein Center, Institute for Theoretical Physics, University of Bern, 3012 Bern, Switzerland}
\affiliation[d]{Fermi National Accelerator Laboratory, Batavia, IL 60510, USA}

\emailAdd{yin01@mit.edu}

\abstract{Complex contour deformations of the path integral have been demonstrated to significantly improve the signal-to-noise ratio of observables in previous studies of two-dimensional gauge theories with open boundary conditions. In this work, new developments based on gauge fixing and a neural network definition of the deformation are introduced, which enable an effective application to theories in higher dimensions and with generic boundary conditions. Improvements of the signal-to-noise ratio by up to three orders of magnitude for Wilson loop measurements are shown in $SU(2)$ lattice gauge theory in three spacetime dimensions.}

\FullConference{%
The 40th International Symposium on Lattice Field Theory,\\
31st July - 4th Aug, 2023,\\
Batavia, Illinois, USA
}

\begin{document}
\maketitle
	
\section{Introduction}
The problem of statistical noise is ubiquitous in lattice field theory calculations, which employ Monte Carlo sampling to estimate field theory expectation values.
Euclidean correlation functions $C(\tau)$ parameterized by imaginary-time separation $\tau$ are generically affected by an exponential decay of the signal-to-noise (StN) ratio with increasing $\tau$~\cite{Parisi:1983ae,Lepage:1989hd},
\begin{equation}
    \mathrm{StN}[C(\tau)] \equiv \left| \left< C(\tau) \right> \right| / \sqrt{\mathrm{Var}[C(\tau)]} \sim \exp({-\textrm{const.} \times \tau}).
\end{equation}
This severe \emph{signal-to-noise problem}
crucially limits the values of $\tau$ which provide meaningful information at a given level of precision.
The signal-to-noise problem also affects correlation functions parameterized by multiple distance scales. 
In lattice gauge theories, the expectation values of rectangular Wilson loops $W_{r \times \tau}$ of dimension $r\times \tau$ can be interpreted as correlation functions of a static particle/anti-particle pair separated by a spatial distance $r$ and Euclidean time $\tau$. A hallmark of confinement in pure gauge theories is the area-law scaling of Wilson loop expectation values,
$
    \left< W_{r \times \tau} \right> \sim e^{-\sigma A}
$,
where
$A = r \times \tau$ and
$\sigma$ is the confining string tension~\cite{Fradkin:1978th}. The signal-to-noise problem can be shown to be maximally severe in this case, with $\mathrm{StN} \sim e^{-\sigma A}$, making this an interesting test bed for approaches to mitigate the signal-to-noise problem in gauge theories.

Recently, \cite{Detmold:2020ncp,Detmold:2021ulb} introduced an approach to signal-to-noise improvement based on \emph{complex contour deformations} of the path integral, which can be used to define families of observables with identical expectation values but generally distinct variance properties.
Similar approaches have been applied to solving sign problems introduced by a complex action~\cite{Cristoforetti:2012su,Cristoforetti:2013wha,Aarts:2013fpa,Alexandru:2015sua,Alexandru:2015xva,Alexandru:2016gsd,Alexandru:2017czx,Alexandru:2017lqr,Alexandru:2018fqp,Alexandru:2020wrj,Mori:2017nwj}, for example arising from a non-zero chemical potential or a real-time path integral; see \cite{Alexandru:2020wrj} for a comprehensive review. Optimizing the choice of deformation was shown in \cite{Detmold:2020ncp,Detmold:2021ulb} to exponentially reduce the noise of the Euclidean-time correlation function of Wilson loops in 2D $U(1)$ and $SU(N)$ lattice gauge theories. However, a key simplification in these results was to first change integration variables in the path integral from gauge links to (untraced) plaquettes as allowed by the 2D geometry with open boundary conditions.

In this work, we introduce a new approach to defining contour deformations of the path integral for $SU(N)$ lattice gauge theory  which is applicable to gauge theories defined in arbitrary spacetime dimensions. In contrast to previous work, these contour deformations are defined to act directly on link degrees of freedom and are therefore applicable to theories defined with generic boundary conditions, including those with periodic boundaries. We introduce and demonstrate the approach in the context of an $SU(2)$ pure gauge theory in 3D, where it is possible to exponentially improve the signal-to-noise ratio of Wilson loops over a wide range of areas. We also investigate the lattice volume dependence of these results. Finally, we conclude by discussing the straightforward generalization to $SU(N)$ and higher dimensions and giving an outlook for this work.

\section{Path integral deformations for signal-to-noise improvement}
We start from the Euclidean path integral definition of the expectation value of a quantum operator $\mathcal{O}$ in a lattice gauge theory,
\begin{equation} \label{eq:path-integral}
    \left< \mathcal{O} \right> \equiv \frac{1}{Z}\int d[U] \mathcal{O}[U] e^{-S[U]},
    \quad
    Z \equiv \int d[U] e^{-S[U]},
\end{equation}
where $\int d[U]$ indicates integration with respect to the Haar measure of gauge field configurations, $S[U]$ is a discretized action, and $\mathcal{O}[U]$ is the representation of the operator acting on the field configurations.
In this work we assume $S[U] \in \mathbb{R}$, in which case equation~\eqref{eq:path-integral} is typically estimated by a statistical expectation value, $\left< \mathcal{O} \right> = \mathbb{E}_S[\mathcal{O}[U]]$,
where $\mathbb{E}_S[\cdot]$ expresses the statistical mean over field configurations $U$ sampled according to the probability measure $d[U] e^{-S[U]}$.

When the Boltzmann weight $\exp(-S[U])$ and operator $\mathcal{O}[U]$ are complex analytic (``holomorphic''), the path integral can be \emph{deformed} continuously into the complexified domain of gauge field configurations without modifying its value, as a result of Cauchy's theorem~\cite{Witten:2010cx,Cristoforetti:2012su}.
For continuous gauge groups, the complexified space can be straightforwardly identified by complexifying the associated Lie algebra; for example, $SU(N)$ is complexified to $SL(N,\mathbb{C})$, the group of $N\times N$ matrices with unit determinant.
Typically, the weight $\exp(-S[U])$ and operators $\mathcal{O}[U]$ are initially defined only on the original (real) domain of integration, but these can be analytically continued to the complexified domain by replacing instances of the Hermitian conjugate links $U^\dag_\mu(x)$ with inverses $U^{-1}_\mu(x)$.

Following \cite{Detmold:2020ncp,Detmold:2021ulb}, we consider a deformation of the original domain of the integral in equation~\eqref{eq:path-integral} to a new domain $\widetilde{\mathcal{M}}$, abstractly written as
\begin{equation}
    \left< \mathcal{O} \right> = \frac{1}{Z} \int_{\widetilde{\mathcal{M}}} d[\tilde{U}] \mathcal{O}[\tilde{U}] e^{-S[\tilde{U}]},
\end{equation}
where we leave the evaluation of $Z$ unchanged. In practice, the integration over the new manifold $\widetilde{\mathcal{M}}$ is defined by an injective map $\tilde{U}(U)$, such that we can write the result of the deformation as
\begin{equation} \label{eq:path-integral-deformed}
\begin{aligned}
    \left< \mathcal{O} \right> &= \frac{1}{Z} \int d[U] J[U] \mathcal{O}[\tilde{U}(U)] e^{-S[\tilde{U}(U)]} \\
    &= \mathbb{E}_S \left[ J[U] \mathcal{O}[\tilde{U}(U)] e^{-S[\tilde{U}(U)]+S[U]} \right] \equiv \mathbb{E}_S \left[ \mathcal{Q}[U]\right],
\end{aligned}
\end{equation}
where $J[U] = \det (\partial \tilde{U} / \partial U)$ is the Jacobian of the map. Because the new observable $\mathcal{Q}[U] = J[U] \mathcal{O}[\tilde{U}(U)] e^{-S[\tilde{U}(U)]+S[U]}$ has identical expectation value to $\mathcal{O}$, one can choose to estimate it instead of $\mathcal{O}[U]$. It will generally have different variance properties from $\mathcal{O}[U]$, and by optimizing the choice of deformation $\tilde{U}(U)$ this variance can be minimized.

\section{Contour deformation of \texorpdfstring{$SU(2)$}{SU(2)} Wilson loops}
We next focus on the signal-to-noise problem of Wilson loops of various sizes in $SU(2)$ pure gauge theory in three dimensions, introducing our choice of deformations $\tilde{U}(U)$ in this context.

\subsection{Complexification and analytic continuation}
We consider the Wilson gauge action, which is given by
\begin{equation}
\begin{aligned}
    S[U] &= - \frac{\beta}{2N_c}\sum_{x, \mu<\nu} \Tr\Big(P_{\mu\nu}(x) +  P^{-1}_{\mu\nu}(x)\Big), \\
    P_{\mu\nu}(x) &= U_\mu(x)U_\nu(x+\hat{\mu})U^{-1}_\mu(x+\hat{\nu})U^{-1}_{\nu}(x).
    \label{eq:S}
\end{aligned}
\end{equation}
Here, $\beta$ is the inverse coupling constant and $P_{\mu\nu}$ is the oriented plaquette located at position $x$. For the $SU(2)$ gauge group, $N_c=2$ and $U_{\mu}(x), P_{\mu\nu}(x) \in SU(2)$.
We have analytically continued the action in the usual way by using inverses $P^{-1}_{\mu\nu}(x)$ and $U^{-1}_{\mu}(x)$ in equation~\eqref{eq:S}.

In this work, we adopt the Bronzan parametrization of $SU(2)$ matrices~\cite{Bronzan:1988wa}
\begin{align} \label{eq:bronzan-param}
    \begin{split}
    U_{\mu}(x) 
    = U\big(\theta_\mu(x), \phi_{1,\mu}(x), \phi_{2,\mu}(x)\big)
    &\equiv
    \begin{pmatrix}
        \sin\big(\theta_\mu(x)\big)e^{i\phi_{1,\mu}(x)}
        &\cos\big(\theta_\mu(x)\big)e^{i\phi_{2,\mu}(x)}\\
        -\cos\big(\theta_\mu(x)\big)e^{-i\phi_{2,\mu}(x)} &
        \sin\big(\theta_\mu(x)\big)e^{-i\phi_{1,\mu}(x)}
    \end{pmatrix},
    \end{split}
\end{align}
where $\theta_\mu(x)\in[0, \pi/2]$ is the polar angle, and $\phi_{1,\mu}(x), \phi_{2,\mu}(x) \in[0, 2\pi)$ are periodic azimuthal angles with periods of $2\pi$. The complexified domain $SL(2,\mathbb{C})$ is given by considering $\theta_\mu(x), \phi_{i,\mu}(x) \in \mathbb{C}$. For general $U_\mu(x) \in SL(2,\mathbb{C})$, the inverse is equal to $U^{-1}_\mu(x) = U(\theta_\mu(x), -\phi_{1,\mu}(x), -\phi_{2,\mu}(x))^T$. In terms of these parameters, the path integral in equation~\eqref{eq:path-integral} can be written as
\begin{align}
    \begin{split}
        \langle O \rangle 
        &= \frac{1}{Z}\int
        \Big(\prod_{x,\mu}d\theta_\mu(x)d\phi_{1,\mu}(x)d\phi_{2,\mu}(x) h(\theta_\mu(x))\Big) 
        e^{-S[U(\Theta)]}
        O\big[U(\Theta)\big],
    \end{split}
    \label{eq:integral_param}
\end{align}
where $\Theta$ indicates the set of all angular variables, $U(\Theta)$ is the set of gauge links parametrized by those variables, and $h_\mu(x) = 1/(4\pi^2)\sin(2\theta_\mu(x))$ is the normalized $SU(2)$ Haar measure.

In \cite{Detmold:2021ulb}, an equivalent parametrization of $SU(2)$ matrices based on Euler angles was also considered and could serve as another starting point for deformation.
Though equivalent, the choice of parameterization can significantly affect whether specific deformations are simple or complicated to write. For the simple class of deformations discussed next, this class defined using the Bronzan parametrization contains all analogous deformations defined using Euler angles, motivating the use of the Bronzan parametrization here.

\subsection{Constant deformations}
To feasibly explore the effects of contour deformations on the signal-to-noise problem, we drastically reduce the space of possible deformations $\tilde{U}(U)$ by considering only variable-independent deformations, or \emph{constant deformations}, which have been used successfully in~\cite{Detmold:2020ncp,Detmold:2021ulb} to reduce Wilson loops variances. In the Bronzan parametrization, this is defined by deforming each $\theta_\mu(x)$, $\phi_{1,\mu}(x)$, and $\phi_{2,\mu}(x)$ by constant shifts into the imaginary direction. Because the $\theta_\mu(x)$ are not periodic, deformations must fix the endpoints  at $0$ and $\pi/2$, which leaves no non-trivial constant shifts for these variables. On the other hand, $\phi_{1,\mu}(x)$ and $\phi_{2,\mu}(x)$ are periodic variables and their contour of integration can safely be shifted, giving in general the deformation map
\begin{align}
    \tilde{U}\big(\theta_\mu(x), \phi_{1,\mu}(x), \phi_{2,\mu}(x)\big) = U\big(\theta_\mu(x), \tilde{\phi}_{1,\mu}(x), \tilde{\phi}_{2,\mu}(x)\big),
\end{align}
where
\begin{align}
    \tilde{\phi}_{i,\mu}(x) = 
    \phi_{i,\mu}(x) + i\Delta_{i, \mu}(x),~(i =1,2),
\end{align}
in terms of the shift fields, $\Delta_{i,\mu}(x)\in \mathbb{R}$.

Because $\Delta_{i,\mu}(x)$ does not depend on the explicit values of $\theta_\mu(x)$ and $\phi_{i,\mu}(x)$, the Jacobian $J[U] = \det (\partial \tilde{U} / \partial U)$ of this transformation is trivial. The normalized Haar measure $h_\mu(x)$ is also not changed by this deformation. We can thus write the new observable
\begin{equation}
  \mathcal{Q}[\Theta] \equiv \mathcal{O}[U(\tilde{\Theta})] e^{-S[U(\tilde{\Theta})]+S[U(\Theta)]},
  \label{eq:constant-deformed-observable}
\end{equation}
where $\tilde{\Theta}$ is the complexified set of all angles, and
as before we have $\left< \mathcal{O} \right> = \mathbb{E}_S \left[ \mathcal{Q}[\Theta] \right]$.

\subsection{Maximal tree gauge fixing}
Unsurprisingly, minimizing the variance of $\mathcal{Q}$ by optimizing the deformation parameters $\Delta_{i,\mu}(x)$ leads to negligible improvement to the signal-to-noise ratio.
Intuitively, constant deformations are restricted in this case to independently deform gauge links, which do not independently encode gauge invariant physical quantities.
Instead, inspired by the previous construction of constant deformations with open boundary conditions in \cite{Detmold:2021ulb}, we first restrict the path integral by fixing to a maximal tree gauge and then deform the remaining degrees of freedom in this formulation. 

Even with a tree-based gauge fixing scheme such as a maximal tree gauge, there is significant choice as to how the fixed links are arranged. We have found that these choices can drastically affect the signal-to-noise improvements which we have been able to achieve. After some exploration, the most significant reduction in variance was observed by choosing the maximal tree gauge defined by 
\begin{align}
    \begin{split}
        &U_0(x_0 < L_0 - 1, x_1=0, x_2=0) = \mathbb{I},\\
        &U_1(x_0, x_1 < L_1 - 1, x_2=0) = \mathbb{I},\\
        &U_2(x_0, x_1, x_2 < L_2 - 1) = \mathbb{I}.
    \end{split}
    \label{eq:gauge-fixed}
\end{align}
Here ${(L_0, L_1, L_2)}$ are the lattice dimensions in lattice units and ${x = (x_0, x_1, x_2)}$ is the lattice coordinate with ${x_i \in \{0,1,\cdots,L_i-1\}}~(i=0,1,2)$.
To obtain the most significant reduction in variance, we also chose the arrangement of the Wilson loop observable to lie in the $(0,1)$-plane with the lower left corner located at the origin ${x = (x_0=0, x_1=0, x_2=0)}$.

\subsection{Deformation parameters from U-nets}
\label{sec:unet}
For small lattices, for example with dimensions $8^3$, optimizing $\Delta_{i,\mu}(x)$ directly on gauge-fixed configurations straightforwardly provides orders-of-magnitude improvement of Wilson loop signal-to-noise ratios. However, if the lattice dimension is taken large while statistics are held fixed, the abundance of parameters $\Delta_{i,\mu}(x)$ quickly leads to an overfitting regime during optimization.
We expect on physical grounds that the reduction in variance achieved for Wilson loops of identical size should be nearly equivalent across lattices with dimensions sufficiently larger than the size of the loop (when $\beta$ is held fixed), but these expected variance improvements are not reached when overfitting is encountered on larger lattices.

\begin{figure}
    \begin{center}
    \includegraphics[scale=0.9]{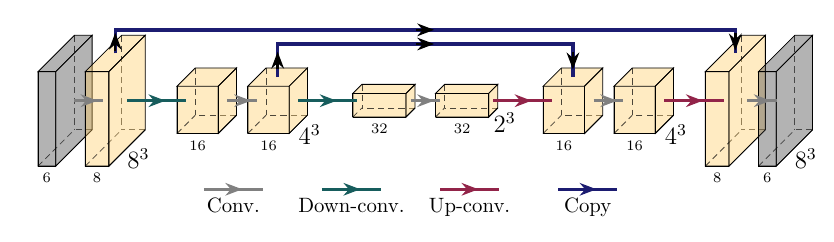}
    \caption{
    U-net architecture used to compute constant shift deformation parameters (gray output) from the gauge fixing and Wilson loop masks (gray input). The architecture is defined using a combination of convolutions (gray arrows), down-convolutions (green arrows), up-convolutions (red arrows), and copy-and-add operations (blue arrows). Each box depicts the number of channels (width) and the 3D lattice geometry (height and depth, third dimension suppressed) at each stage of the network. The 6 input channels correspond to gauge-fixing masks and the Wilson loop masks for the three orientations of links, while the six output channels correspond to the components of $\Delta_{i,\mu}(x)$.
    }
    \label{fig:u-net}
    \end{center}
\end{figure}

\begin{figure}
    \begin{center}
    \includegraphics[scale=0.9]{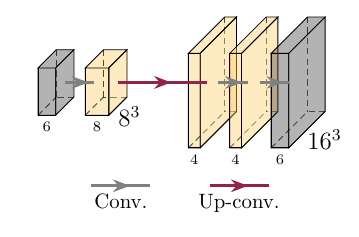}
    \caption{The decoder architecture that takes $\Delta_{i,\mu}(x)$ for a Wilson loop of size $r \times \tau$ in an $8^3$ lattice geometry as the input and outputs $\Delta_{i,\mu}(x)$ for a Wilson loop of size $(2 r)\times (2 \tau)$ in a $16^3$ lattice geometry. The meanings of the operations and boxes are the same as in figure~\ref{fig:u-net}.}
    \label{fig:transfer}
    \end{center}
\end{figure}

To overcome this training issue, we reparametrize the shift field, $\Delta_{i,\mu}(x)$, as
\begin{align}
    \Delta_{i,\mu}(x) = f_{i,\mu}(x;w),
    \label{eq:u-net}
\end{align}
where $f_{i,\mu}(x;w)$ is the output of a neural network and $w$ is the set of weights defining the neural network, which implicitly parameterize this output. This is a new parameterization of approximately the same space of deformations, albeit one where the parameters $w$ at some finite neural network complexity are less prone to overfitting. To produce $f_{i,\mu}(x;w)$, we use the {\it U-net architecture}~\cite{ronneberger2015unet} applied to a constant input given by a collection of Boolean fields specifying the gauge-fixed links and the links making up the Wilson loop. Figure~\ref{fig:u-net} shows an example of the U-net for $8^3$ lattices.

For a generic lattice geometry, the U-net is constructed so that each coarsening step (``down-convolution'') reduces the lattice dimensions by a factor of two along each axis while increasing the number of channel dimensions by a factor of two; on the other hand, the prolongation steps (``up-convolution'') double the lattice dimensions along each axis while reducing the channel dimension by a factor of two. In this work, we always coarsen until the lattice dimensions are $2^3$ at the coarsest level before prolongation. Because the gauge-fixed links cannot be deformed, we multiply by a mask as a final step to enforce $f_{i,\mu}(x; w) = 0$ for all $\mu,x$ where $U_\mu(x)$ are gauge-fixed.

The use of the U-net in training not only allows us to reach much smaller variances on lattice sizes $\gtrapprox16^3$, but also enables much faster training on smaller lattices by reducing the total number of updating steps to the weights. For these reasons, all results in this work are based on this U-net parametrization of the shift field.

\subsection{Decoder network and transfer learning}
\label{sec:transfer}
The U-net parametrization by itself is sufficient to parameterize the shift field at any given lattice volume and Wilson loop geometry. However, as a practical improvement, we additionally introduce and study the use of a \emph{decoder network} that uses convolutions to transform a shift field from a smaller volume to a larger volume for lattices with identical $\beta$ and gauge fixing schemes. Figure~\ref{fig:transfer} shows an example of the convolutional decoder architecture that we use to transform shift fields between $8^3$ and $16^3$ lattice volumes.

To train the network to transform between a smaller and larger lattice geometry differing by a factor of two in each dimension, we first train a U-net to produce a shift field $\Delta_{i,\mu}(x)$ for a Wilson loop size of $r \times \tau$ on the smaller lattice volume. The decoder takes this shift field as the input and outputs a new shift field which is used to deform the path integral for a Wilson loop size of $(2 r)\times (2 \tau)$, with the variance of this larger loop used as a target for optimization. 

Comparing deformations produced from a U-net targeting the final lattice geometry versus using the U-net output at smaller volume together with a decoder network, we find that both reach the same variance improvement once properly trained. However, once the decoder network is trained to transform from $8^3$ to $16^3$ lattice volumes, we can fine-tune the same network by a smaller number of optimization steps to transform the shift field on the $16^3$ lattices to a shift field suitable for $32^3$ lattices. This transfer learning scheme is crucial for training on lattice volumes $\gtrapprox32^3$ where the training costs needed for directly training a U-net become large.

\section{Training details and results}
In this work, we present results for lattice theories with fixed $\beta = 3.75$, periodic boundary conditions, and three choices of lattice sizes, $8^3$, $16^3$, and $32^3$. The bare coupling corresponds to a string tension of $a^2 \sigma \approx (0.4)^2$ in lattice units~\cite{Athenodorou:2016ebg}. For the three volumes, we respectively generate $1.5\times 10^5$, $1.1\times 10^5$, and $2.4\times 10^4$ configurations for training and $500$ configurations for testing. No significant autocorrelations of plaquette values are observed within these configurations.

For the $8^3$ ensemble, we find the optimal constant deformation by directly training U-nets as described in section~\ref{sec:unet}; for the $16^3$ ensemble, we use the decoder architecture shown in figure~\ref{fig:u-net}. We also compared directly training a U-net on the $16^3$ ensemble, with those two methods giving the same performance. For the $32^3$ ensemble,
we reuse the decoder network previously trained for the $16^3$ ensemble and fine-tune it with fewer than $10^3$ steps of weight updates, 
a much lesser amount of additional training than directly training the shift fields.

To avoid numerical stability issues, we train in all cases by minimizing the logarithm of the variance of the targeted Wilson loop for each geometry using the numerically stable LogSumExp function. We also use the trick introduced in \cite{Detmold:2021ulb} of measuring  and optimizing only the $(1,1)$ component of the untraced Wilson loop.
All training is done on a single compute node with eight NVIDIA A100 GPUs. We use mini-batch sizes of $32$ configurations per GPU with the default settings of the Adam optimizer~\cite{DBLP:journals/corr/KingmaB14}. Training for each Wilson loop size takes at most a few node-hours.

\begin{figure*}
    \begin{center}\includegraphics[width=0.7\textwidth]{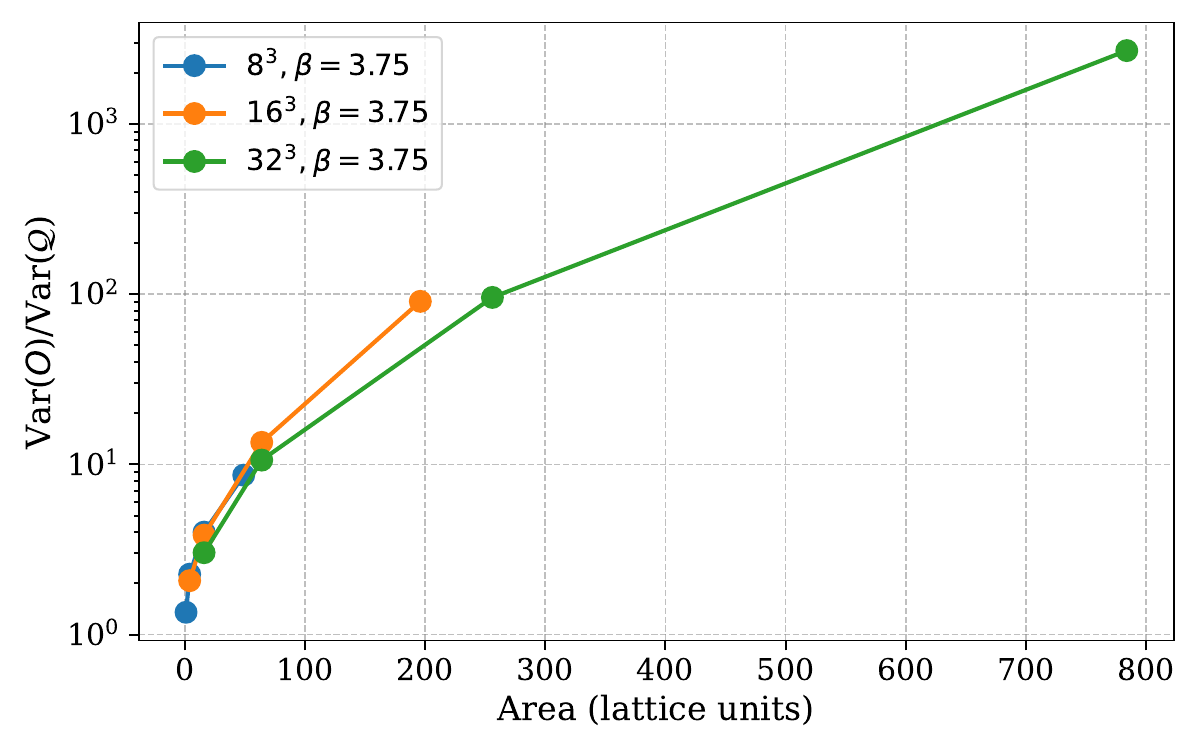}
    \caption{Ratios of variances between the original ($O$) and improved ($Q$) Wilson loop observables of various areas for $8^3$, $16^3$, and $32^3$ ensembles.}
    \label{fig:results}
    \end{center}
\end{figure*}

The resulting improvement factors in the Wilson loop variance are shown in figure~\ref{fig:results}.
Up to three orders of magnitude of variance reduction is obtained at the largest Wilson loop studied. As the area is increased, we find increasing improvements.
The reduction in variance obtained using the techniques discussed above is also nearly independent of the total lattice volume, depending much more strongly on the Wilson loop geometry.

\section{Outlook and conclusion}
In this work, we showed for the first time how to use constant contour deformations to mitigate the signal-to-noise problem of Wilson loops for three dimensional $SU(2)$ gauge theory with periodic boundary conditions. To achieve this result, we used a maximal tree gauge fixing scheme (equation~\eqref{eq:gauge-fixed}), a U-net parametrization of the deformation (equation~\eqref{eq:u-net} and figure~\ref{fig:u-net}), and decoder networks with transfer learning (section~\ref{sec:transfer} and figure~\ref{fig:transfer}).

Our approach of constant contour deformations parametrized by a U-net can easily be generalized to 4D by using four-dimensional convolutions and a four-dimensional maximal tree gauge. These deformations are also applicable to other choices of $N_c$, including the $SU(3)$ gauge group, by working with the corresponding Bronzan angular parametrizations~\cite{Bronzan:1988wa}. We are currently working to apply the method to four dimensional $SU(2)$ and $SU(3)$ gauge theories.

Within the framework of the constant deformation presented in this work, many avenues are left to be explored that may lead to better performance. Most importantly, we have observed that the choice of the gauge fixing scheme heavily influences the final variances of improved observables. In addition, the maximal tree gauge fixing scheme we use here completely obscures the translation symmetry of the lattice, as it must be combined with gauge transformations to maintain the gauge fixing scheme. This means that performing volume averaging that is commonly used for variance reduction is expensive since each translation requires additional gauge fixing. Other gauge fixing schemes, such as partial subsets of a maximal tree gauge, could simplify the process of (partial) volume averaging and may yield similar or better variance improvements.
Finally, field-independent deformations represent a small class of all possible deformations. Exploring contour deformation beyond this simple choice, such as the Fourier series expansion presented in~\cite{Detmold:2021ulb}, may further boost the performance and eliminate the needs of gauge fixing completely.

\bibliographystyle{utphys}
\bibliography{bib}

\providecommand{\href}[2]{#2}\begingroup\raggedright\begin{thebibliography}{10}

\bibitem{Parisi:1983ae}
G.~Parisi \href{http://dx.doi.org/10.1016/0370-1573(84)90081-4}{{\em Phys.
  Rept.} {\bfseries 103} (1984) 203--211}.

\bibitem{Lepage:1989hd}
G.~P. Lepage in {\em {Theoretical Advanced Study Institute in Elementary
  Particle Physics}}.
\newblock 6, 1989.

\bibitem{Fradkin:1978th}
E.~H. Fradkin and L.~Susskind
  \href{http://dx.doi.org/10.1103/PhysRevD.17.2637}{{\em Phys. Rev. D}
  {\bfseries 17} (1978) 2637}.

\bibitem{Detmold:2020ncp}
W.~Detmold, G.~Kanwar, M.~L. Wagman, and N.~C. Warrington
  \href{http://dx.doi.org/10.1103/PhysRevD.102.014514}{{\em Phys. Rev. D}
  {\bfseries 102} no.~1, (2020) 014514},
  \href{http://arxiv.org/abs/2003.05914}{{\ttfamily arXiv:2003.05914
  [hep-lat]}}.

\bibitem{Detmold:2021ulb}
W.~Detmold, G.~Kanwar, H.~Lamm, M.~L. Wagman, and N.~C. Warrington
  \href{http://dx.doi.org/10.1103/PhysRevD.103.094517}{{\em Phys. Rev. D}
  {\bfseries 103} no.~9, (2021) 094517},
  \href{http://arxiv.org/abs/2101.12668}{{\ttfamily arXiv:2101.12668
  [hep-lat]}}.

\bibitem{Cristoforetti:2012su}
{AuroraScience} Collaboration, M.~Cristoforetti, F.~Di~Renzo, and L.~Scorzato
  \href{http://dx.doi.org/10.1103/PhysRevD.86.074506}{{\em Phys. Rev. D}
  {\bfseries 86} (2012) 074506},
  \href{http://arxiv.org/abs/1205.3996}{{\ttfamily arXiv:1205.3996 [hep-lat]}}.

\bibitem{Cristoforetti:2013wha}
M.~Cristoforetti, F.~Di~Renzo, A.~Mukherjee, and L.~Scorzato
  \href{http://dx.doi.org/10.1103/PhysRevD.88.051501}{{\em Phys. Rev. D}
  {\bfseries 88} no.~5, (2013) 051501},
  \href{http://arxiv.org/abs/1303.7204}{{\ttfamily arXiv:1303.7204 [hep-lat]}}.

\bibitem{Aarts:2013fpa}
G.~Aarts \href{http://dx.doi.org/10.1103/PhysRevD.88.094501}{{\em Phys. Rev. D}
  {\bfseries 88} no.~9, (2013) 094501},
  \href{http://arxiv.org/abs/1308.4811}{{\ttfamily arXiv:1308.4811 [hep-lat]}}.

\bibitem{Alexandru:2015sua}
A.~Alexandru, G.~Basar, P.~F. Bedaque, G.~W. Ridgway, and N.~C. Warrington
  \href{http://dx.doi.org/10.1007/JHEP05(2016)053}{{\em JHEP} {\bfseries 05}
  (2016) 053}, \href{http://arxiv.org/abs/1512.08764}{{\ttfamily
  arXiv:1512.08764 [hep-lat]}}.

\bibitem{Alexandru:2015xva}
A.~Alexandru, G.~Basar, and P.~Bedaque
  \href{http://dx.doi.org/10.1103/PhysRevD.93.014504}{{\em Phys. Rev. D}
  {\bfseries 93} no.~1, (2016) 014504},
  \href{http://arxiv.org/abs/1510.03258}{{\ttfamily arXiv:1510.03258
  [hep-lat]}}.

\bibitem{Alexandru:2016gsd}
A.~Alexandru, G.~Basar, P.~F. Bedaque, S.~Vartak, and N.~C. Warrington
  \href{http://dx.doi.org/10.1103/PhysRevLett.117.081602}{{\em Phys. Rev.
  Lett.} {\bfseries 117} no.~8, (2016) 081602},
  \href{http://arxiv.org/abs/1605.08040}{{\ttfamily arXiv:1605.08040
  [hep-lat]}}.

\bibitem{Alexandru:2017czx}
A.~Alexandru, P.~F. Bedaque, H.~Lamm, and S.~Lawrence
  \href{http://dx.doi.org/10.1103/PhysRevD.96.094505}{{\em Phys. Rev. D}
  {\bfseries 96} no.~9, (2017) 094505},
  \href{http://arxiv.org/abs/1709.01971}{{\ttfamily arXiv:1709.01971
  [hep-lat]}}.

\bibitem{Alexandru:2017lqr}
A.~Alexandru, G.~Basar, P.~F. Bedaque, and G.~W. Ridgway
  \href{http://dx.doi.org/10.1103/PhysRevD.95.114501}{{\em Phys. Rev. D}
  {\bfseries 95} no.~11, (2017) 114501},
  \href{http://arxiv.org/abs/1704.06404}{{\ttfamily arXiv:1704.06404
  [hep-lat]}}.

\bibitem{Alexandru:2018fqp}
A.~Alexandru, P.~F. Bedaque, H.~Lamm, and S.~Lawrence
  \href{http://dx.doi.org/10.1103/PhysRevD.97.094510}{{\em Phys. Rev. D}
  {\bfseries 97} no.~9, (2018) 094510},
  \href{http://arxiv.org/abs/1804.00697}{{\ttfamily arXiv:1804.00697
  [hep-lat]}}.

\bibitem{Alexandru:2020wrj}
A.~Alexandru, G.~Basar, P.~F. Bedaque, and N.~C. Warrington
  \href{http://dx.doi.org/10.1103/RevModPhys.94.015006}{{\em Rev. Mod. Phys.}
  {\bfseries 94} no.~1, (2022) 015006},
  \href{http://arxiv.org/abs/2007.05436}{{\ttfamily arXiv:2007.05436
  [hep-lat]}}.

\bibitem{Mori:2017nwj}
Y.~Mori, K.~Kashiwa, and A.~Ohnishi
  \href{http://dx.doi.org/10.1093/ptep/ptx191}{{\em PTEP} {\bfseries 2018}
  no.~2, (2018) 023B04}, \href{http://arxiv.org/abs/1709.03208}{{\ttfamily
  arXiv:1709.03208 [hep-lat]}}.

\bibitem{Witten:2010cx}
E.~Witten {\em AMS/IP Stud. Adv. Math.} {\bfseries 50} (2011) 347--446,
  \href{http://arxiv.org/abs/1001.2933}{{\ttfamily arXiv:1001.2933 [hep-th]}}.

\bibitem{Bronzan:1988wa}
J.~B. Bronzan \href{http://dx.doi.org/10.1103/PhysRevD.38.1994}{{\em Phys. Rev.
  D} {\bfseries 38} (1988) 1994}.

\bibitem{ronneberger2015unet}
O.~Ronneberger, P.~Fischer, and T.~Brox in {\em MICCAI 2015: 18th International
  Conference}, pp.~234--241, Springer.
\newblock 2015.
\newblock \href{http://arxiv.org/abs/1505.04597}{{\ttfamily arXiv:1505.04597
  [cs.CV]}}.

\bibitem{Athenodorou:2016ebg}
A.~Athenodorou and M.~Teper
  \href{http://dx.doi.org/10.1007/JHEP02(2017)015}{{\em JHEP} {\bfseries 02}
  (2017) 015}, \href{http://arxiv.org/abs/1609.03873}{{\ttfamily
  arXiv:1609.03873 [hep-lat]}}.

\bibitem{DBLP:journals/corr/KingmaB14}
D.~P. Kingma and J.~Ba in {\em {ICLR} 2015}, Y.~Bengio and Y.~LeCun, eds.
\newblock 2015.
\newblock \url{http://arxiv.org/abs/1412.6980}.

\end{thebibliography}\endgroup
\end{document}